\documentclass[journal,comsoc]{IEEEtran}

\usepackage{standalone}
\usepackage[T1]{fontenc}
\usepackage{algorithm}
\usepackage[normalem]{ulem}
\usepackage{tikz,tkz-euclide}
    \usetikzlibrary{decorations}
    \usetikzlibrary{decorations.pathreplacing}
    \usetikzlibrary{shapes,arrows,shadows}
    \usetikzlibrary{calc,positioning}
    \usetikzlibrary{fit,shapes}
    \usetikzlibrary{calc}
\usepackage{balance}
\usepackage{pgfplots,siunitx}
\pgfplotsset{compat=1.14}
\usepackage{cancel}
\usepackage{soul}
\usepackage{amsmath,amssymb,amsthm}
	\theoremstyle{definition}
	
	\newtheorem{example}{Example}
	\newtheorem{remark}{Remark}

\usetikzlibrary{plotmarks,matrix,chains,scopes,fit,calc,shapes,positioning,decorations,intersections,arrows,backgrounds,shadows}
\newlength\FigureHeight
\newlength\FigureWidth
\newcommand{\define}{\stackrel{\Delta}{=}}

\def\Semax{\mathcal{S}_{\emax}^{N,m}}
\def\So{\mathcal{S}_{\circ}}
\def\calA{\mathcal{A}}

\def\calS{\mathcal{S}}
\def\calC{\mathcal{C}}

\def\calX{\mathcal{X}}

\def\rloss{R_{\text{loss}}}

\def\E{ E_{\text{av}}}

\def\emax{E_{\max}}

\def\Gs{G_{\text{s}}}

\def\rloss{R_{\text{loss}}}

\def\ro{r_{\circ}}

\def\snr{\text{SNR}}

\def\ent{\mathbb{H}}

\def\rbmd{R_{\text{BMD}}}
\def\capc{C}
\def\delsnr{\Delta\text{SNR}}
\colorlet{shapercolor}{green!20!white}
\colorlet{bshapercolor}{cyan!20!white}
\colorlet{feccolor}{blue!20!white}
\colorlet{mapcolor}{red!20!white}
\colorlet{chancolor}{black!10!white}
\def\lc{\left\lceil}   
\def\rc{\right\rceil}
\def\lf{\left\lfloor}   
\def\rf{\right\rfloor}

\def\calC{\mathcal{C}}
\def\calS{\mathcal{S}}

\def\calA{\mathcal{A}}
\def\calX{\mathcal{X}}

\def\ent{\mathbb{H}}

\def\Emax{E_{\text{max}}}

\begin{document}
\title{Enumerative Sphere Shaping for Wireless Communications with Short Packets}

\author{Yunus~Can~G\"{u}ltekin,~\IEEEmembership{Student Member,~IEEE,}
        Wim J. van Houtum,~\IEEEmembership{Senior Member,~IEEE,}
        Arie Koppelaar,
        Frans M. J. Willems,~\IEEEmembership{Fellow,~IEEE,}
\thanks{Manuscript received March 21, 2019; revised August 13, 2019 and October 21; accepted  October 28, 2019. 
The material in this paper was presented in part at the IEEE International Symposium on Personal, Indoor and Mobile Radio Communications (PIMRC), Montreal, QC, Canada, October 08-13, 2017.}
\thanks{Y. C. G\"{u}ltekin, W. J. van Houtum and F. M. J. Willems are with the Information and Communication Theory Lab, Signal Processing
Systems Group, Department of Electrical Engineering, Eindhoven University
of Technology, Eindhoven 5600 MB, the Netherlands (e-mail: \{y.c.g.gultekin;  w.j.v.houtum; f.m.j.willems\}@tue.nl).}
\thanks{W. J. van Houtum is also with Catena Radio Design, Eindhoven, the Netherlands.}
\thanks{A. Koppelaar is with NXP-Research, Eindhoven, the Netherlands (e-mail: arie.koppelaar@nxp.com).}
\thanks{The work of Y. C. G\"{u}ltekin is supported by TU/e Impuls program, a strategic cooperation between NXP Semiconductors and Eindhoven University of Technology}
}

\maketitle

\begin{abstract}
Probabilistic amplitude shaping (PAS) combines an outer shaping layer with an inner, systematic forward error correction (FEC) layer to close the shaping gap. 
Proposed for PAS, constant composition distribution matching (CCDM) produces amplitude sequences with a fixed empirical distribution. 
We show that CCDM suffers from high rate losses for small block lengths, and we propose to use Enumerative Sphere Shaping (ESS) instead. 
ESS minimizes the rate loss at any block length.
Furthermore, we discuss the computational complexity of ESS and demonstrate that it is significantly smaller than shell mapping (SM), which is another method to perform sphere shaping. 
We then study the choice of design parameters for PAS. 
Following Wachsmann \textit{et al.}, we show that for a given constellation and target rate, there is an optimum balance between the FEC code rate and the entropy of the Maxwell-Boltzmann distribution that minimizes the gap-to-capacity.
Moreover, we demonstrate how to utilize the non-systematic convolutional code from IEEE 802.11 in PAS. 
Simulations over the additive white Gaussian noise (AWGN) and frequency-selective channels exhibit that ESS is up to 1.6 and 0.7 dB more energy-efficient than uniform signaling at block lengths as small as 96 symbols, respectively, with convolutional and low-density parity-check (LDPC) codes. 
\end{abstract}

\begin{IEEEkeywords}
Sphere Shaping, Probabilistic Amplitude Shaping, Amplitude Shift Keying.
\end{IEEEkeywords}

\IEEEpeerreviewmaketitle

\section{Introduction}
\IEEEPARstart{T}{he} maximum transmission rate at which reliable communication is possible over the additive white Gaussian noise (AWGN) channel requires Gaussian signaling~\cite{CoverT2006_ElementsofInfoTheo}.
Gaussian signaling realizes that the logarithm of the probability of transmitting a signal point is negatively proportional to its energy.  
Conversely, uniform signaling implies that all possible signals are transmitted with equal probability.
The increase in required signal-to-noise ratio (SNR) to achieve a rate when applying a uniform signaling strategy instead of a Gaussian one is called the shaping gap and is asymptotically equal to 1.53 dB for large rates.
Equivalently, the shaping gap can be expressed as a decrease in the achievable information rate (AIR) and is equal to 0.255 bits per real dimension asymptotically for large SNRs.

The techniques which are developed to close the shaping gap can be classified into two main groups: {\it Geometric shaping} and {\it probabilistic shaping}.
Geometric shaping employs equiprobable signaling with Gaussian-like distributed signal points~\cite{sun1993}.
{Probabilistic shaping,} as the name implies, describes methods that choose channel inputs (which are usually equidistant) according to a Gaussian-like distribution~\cite{calderbank1990, frank1993}.
In this work, we limit our attention to {probabilistic shaping} and refer to~\cite[Sec. 4.5]{Fischer2002} for a discussion on geometric shaping.
{Furthermore, a detailed discussion on the taxonomy of signal shaping is provided in~\cite[Sec. I]{Gultekin2019Arxiv_Comparison}.}

Regarding {probabilistic shaping,} there exist two fundamental paradigms which are called the direct and the indirect methods by Calderbank and Ozarow in~\cite{calderbank1990}.
The first and information-theoretically elegant approach is to {\it directly} start with a target probability distribution (preferably the one that achieves capacity) over a low-dimensional constellation and endeavor to realize it during transmission.
For this purpose, Maxwell-Boltzmann (MB) distributions are commonly considered since they are the discrete domain counterpart of the Gaussian distributions and maximize the entropy for a fixed average energy~\cite{CoverT2006_ElementsofInfoTheo}.
The other method is to change the bounding geometry of the signal-space which {\it indirectly} induces a non-uniform distribution on lower dimensions.
As for this, an $N$-sphere is the natural selection as it is the most energy-efficient geometry given its volume~\cite{forney1984}.
A concise review of {probabilistic shaping} schemes can be found in~\cite[Sec. II]{bocherer2015}.

Recently, probabilistic amplitude shaping (PAS) is proposed as a solution to combine shaping with channel coding in~\cite{bocherer2015}.
PAS is a reversely concatenated scheme where the shaping operation precedes channel coding at the transmitter side.
This construction is first studied in the context of constrained coding~\cite{bliss1981} and a corresponding soft-decision decoding approach was investigated in \cite{fan1999}.
The strategy of having an outer shaper-deshaper pair has two main advantages:
First, the transmission rate of the PAS scheme can easily be adapted by changing the rate of the amplitude shaping block for a fixed forward error correction (FEC) code.
Second, since the decoding of the FEC code precedes the deshaping operation in PAS, the requirement to apply a soft-output deshaper which may have an extensive complexity is removed.

In the PAS framework, the block that realizes shaping determines the amplitudes of the channel inputs.
Then based on the binary labels of these amplitudes, a systematic FEC encoder selects the signs.
The function of the shaping block is to map uniform information bits to amplitude sequences that satisfy a pre-defined condition in an invertible manner.
Amplitude shaping can be accomplished in two steps. 
First, a shaping set, i.e., the set of amplitude sequences, is constructed such that the desired distribution is obtained (direct method) or the average energy is minimized (indirect method).
The set construction method is called the {\it shaping architecture}, and the performance of the architecture is related to the achieved set size and expressed in rate loss.
Second, an encoding function that maps uniform bit sequences to amplitude sequences from the shaping set is implemented.
This implementation determines the complexity and is called the {\it shaping algorithm}.
The distinction between the shaping architecture and algorithm was also made in~\cite[Sec. I]{Fehenberger2019Arxiv_PASR} {and further discussed in~\cite[Sec. II-D]{Gultekin2019Arxiv_Comparison}}.

{For the initial proposal of PAS~\cite{bocherer2015}, constant composition distribution matching (CCDM) is employed as the amplitude shaping architecture~\cite{ccdm}.
CCDM uses the direct method according to the Calderbank-Ozarow terminology.
As the shaping set, the set of amplitude sequences that have the same composition is considered. 
This composition is selected such that the corresponding distribution is information-theoretically {\it close} to a target distribution.
Targeting MB distributions, CCDM is used in PAS to operate less than 1.1 dB away from the AWGN channel capacity~\cite{bocherer2015}.
As the shaping algorithm, arithmetic coding (AC) is employed to realize CCDM~\cite{ccdm,ramabadran1990}.}

The performance of CCDM improves with increasing block lengths which makes it suitable for applications having long data packets~\cite{ccdm}.
As an example, the second generation Digital Video Broadcasting standard employs low-density parity-check (LDPC) codes of length 64800~bits~\cite{dvbs2}.
This is equivalent to having 21600 amplitudes at the output of the matcher when 8-ary amplitude shift keying (ASK) is used. 
At such large block lengths, the rate losses that CCDM suffers from can be ignored.
However in cases where the output of the shaping block is less than a couple of hundreds symbols long, CCDM experiences significant rate losses~\cite{ccdm}.
As another example, the IEEE 802.11 standard uses channel codewords as small as 648 bits which leads to block lengths of around 200 symbols~\cite[Table 19-15]{80211-2016}. 
At such small block lengths, the rate losses that CCDM encounters makes it less suitable as a shaping technique.

To replace CCDM in the moderate block length regime, Fehenberger \textit{et al.} introduced~in \cite{pbdm} the concept of multiset-partition distribution matching (MPDM).
In short, MPDM builds upon the principle of CCDM and allows multiple amplitude compositions for the sequences that are produced by the matcher.
By selecting a set of compositions such that their average is the target composition, MPDM provides a better rate loss performance than CCDM and is suitable for application for shorter block lengths. 
However, the number compositions can be quite large.

In the current paper, we investigate sphere shaping of multidimensional constellations.
In particular, we are interested in enumerative sphere shaping (ESS) as the amplitude shaping function within the PAS framework.
{ESS uses the indirect method according to the Calderbank-Ozarow terminology, and provides a computationally efficient algorithm to realize sphere shaping.}
Recently, the performance of ESS was numerically and experimentally evaluated for optical communication systems in~\cite{Amari2019_IntroducingESSoptics,Amari2019_ESSreachincrease,Goossens2019_FirstExperimentESS}.
By comparing ESS to CCDM and to sphere shaping algorithms described by Laroia {\it et al.}~\cite{laroia1994}, we will demonstrate that it is a shaping technique with an excellent performance for small to moderate block lengths, and evaluate its performance over the AWGN and frequency selective channels.

Our contribution is threefold.
After providing background information in Sec.~\ref{sec:background}, first we explain the enumerative approach to realize sphere shaping and propose to use it as the shaping procedure in the PAS construction in Sec.~\ref{sec:ess}.
Considering the signal structure, we see that CCDM employs {\it some} of the signal points located on the surface of a multidimensional shell.
MPDM then utilizes sequences from multiple partially filled shells.
On the other hand, sphere shaping is able to make use of all sequences inside a sphere as shown in Fig.~\ref{surface2sphere}.
Thus, it achieves the smallest rate loss possible for a given target rate.
In addition, in Sec.~\ref{sec:implementation}, we compare the enumerative approach with shell mapping (SM) which is an alternative method for realizing sphere shaping and we show that the enumerative approach requires a significantly smaller computational complexity for moderate block lengths.

Second, we study the selection of parameters for the PAS scheme in Sec.~\ref{sec:shapcode}.
Following Wachsmann \textit{et al.}~\cite{wachsmann1999}, we explain how the optimum shaping and coding rates should be selected for a given constellation size and target rate for the AWGN channel.
Furthermore, we do a similar analysis for fading channels {for the first time ---to the best of our knowledge---} and show that it is possible to increase the AIR with limited shaping redundancy.

Subsequently, we explain how a non-systematic code, namely the convolutional code from the IEEE 802.11 standard~\cite{80211-2016} can be used as the inner FEC code of PAS in Sec.~\ref{nsfecpas}.
One of the important properties of PAS is that it combines shaping with existing systematic FEC codes.
We broaden its application area by providing a method to integrate a non-systematic code into it.
Finally, simulation results for ESS over the AWGN and frequency-selective channels are provided in Sec.~\ref{results} before the conclusions.

\section{Background on Amplitude Shaping} \label{sec:background}
\subsection{Channel Capacity and Amplitude Shaping} \label{ssec:capacity}
The time-discrete AWGN channel is modeled at time $n = 1, 2,\cdots, N$ as\footnote{Notation: Capital letters $X$ indicate random variables. Random vectors are denoted by $X^N$. Realizations of random variables and vectors are indicated by $x$ and $x^N$, respectively. 
{Calligraphic letters $\calX$ represent sets.
The Cartesian product of $\calX$ and $\mathcal{Y}$ is indicated as $\calX \times \mathcal{Y}$.}
Boldface capital letters $\boldsymbol{G}$ specify matrices. $P_X(x)$ denotes the probability mass function (PMF) for $X$. Probability density function (PDF) of $Y$ conditioned on $X$ is denoted by $f_{Y|X}(y|x)$.}
\begin{eqnarray}
Y_n = X_n + Z_n,
\end{eqnarray}
where $N$ is the block length in real symbols, $X_n$ is the channel input and $Y_n$ is the output.
Here $Z_n$ is the noise that is independent of $X_n$ and drawn from a zero-mean Gaussian distribution with variance $\sigma^2$. 
The channel inputs are power-constrained, i.e., $\mathbb{E}[X^2] \leq P$, where $\mathbb{E}$ denotes the expectation operator.

Defining the signal-to-noise ratio $\snr = \mathbb{E}[X^2]/\sigma^2$, the capacity of the AWGN channel is $C = \frac{1}{2}\log_2(1+\snr)$ in bits per real dimension (bit/1-D).
This capacity is achieved when $X$ is a zero-mean Gaussian with variance $P$~\cite{CoverT2006_ElementsofInfoTheo}.
The corresponding random coding argument shows that code sequences, drawn from a Gaussian distribution, are likely to lie in an $N$-sphere of squared radius $NP(1+\varepsilon)$ for any $\varepsilon>0$, when $N\rightarrow \infty$. 
Therefore, it makes sense to choose the codewords inside a sphere, or equivalently to use an $N$-sphere as the signal space boundary, to achieve capacity $C$.
Alternatively, the sphere hardening result, see e.g. Wozencraft and Jacobs \cite{wozencraft1965}, shows that practically all codewords are near the surface of the sphere as $N\rightarrow \infty$. 
Therefore, one could argue that codewords chosen from the surface of a sphere would lead to good signal sets. 
In the current paper, we are comparing both approaches, the first approach we refer to as sphere shaping, the second approach {(CCDM)} gives codewords on the surface of a sphere by using constant composition codewords.

Throughout our paper, we consider $2^m$-ASK that results in alphabet $\calX = \left\{\pm 1, \pm 3,\cdots, \pm (2^m-1)\right\}$ for $m \geq 2$.
This alphabet can be factorized as $\calX = \calS \times \calA$ where $\calS = \{-1, 1\}$ and $\calA = \{+1,+3,\cdots, 2^m-1 \}$ are the sign and amplitude alphabets, respectively.
The amplitude shaping method determines which amplitudes are used for the channel input sequences, whereas FEC will be added later to determine the signs of the channel inputs.
The implicit assumption here is that the FEC code is systematic with a rate larger than or equal to $(m-1)/m$, i.e., the FEC encoder adds at most 1 bit of redundancy per ASK symbol.

\subsection{Constant Composition Distribution Matching}
\begin{figure}[t]
\centering
\resizebox{\columnwidth}{!}{\includegraphics{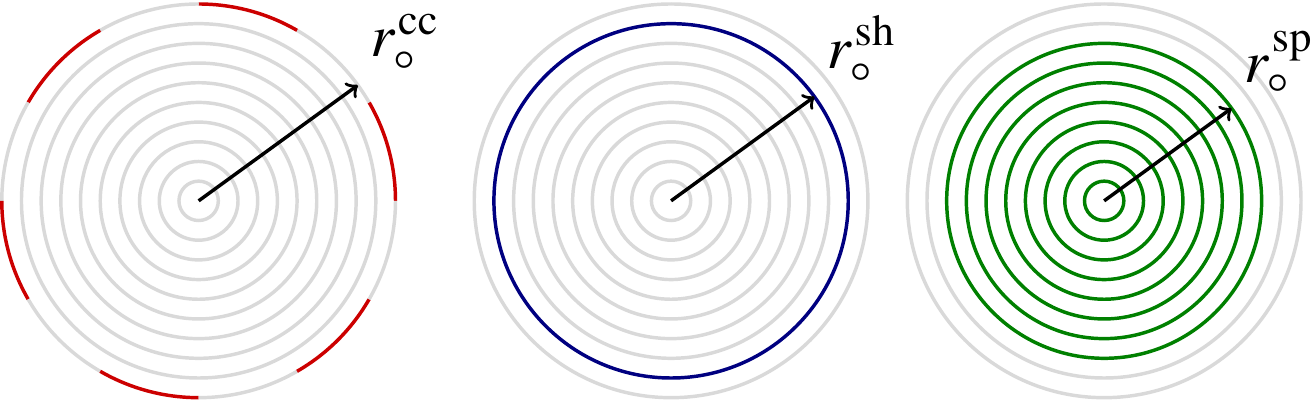}}
\caption{The illustration of the utilized sequences by constant composition (left), single-shell (middle) and sphere codes (right). Each circle depicts an $N$-dimensional shell. Darker portions of the shells represent the sequences on them which are utilized by the corresponding shaping approach. When all include the same number of sequences, radii satisfy $r^{\text{cc}}_{\circ} \geq r^{\text{sh}}_{\circ} \geq r^{\text{sp}}_{\circ}$ at any dimension.}
\label{surface2sphere}
\end{figure}

We will first discuss constant composition codes to realize amplitude shaping (direct method).
The idea is to utilize $N$-sequences having a fixed empirical distribution. 
To this end, a target amplitude composition $\{ \#(a) \approx N P_A(a), a\in\calA\}$ is found~\cite{ccdm}, where $P_A$ denotes the target distribution\footnote{The notation $\#(a)$ is used to indicate the number of occurrences of $a$ in a constant composition $N$-sequence. 
Obviously, $\sum_{a\in\calA} \#(a) = N$.}.
The shaping rate of the constant composition code which consists of all $N$-sequences having composition $\{\#(a), a\in\calA \}$  is
\begin{eqnarray}
R_s = \frac{1}{N} \log_2\left( \frac{N!}{\prod_{a\in \calA}\#(a)!} \right),
\end{eqnarray}
in bits per amplitude (bit/amp.) and converges to $\ent(P_A)$ asymptotically for large $N$.
$\ent(P_A)$ denotes the entropy of $P_A$ in bits.
All sequences in the code have the same energy, and thus, the average sequence energy is $\E = \sum_{a\in\calA} \#(a)a^2$. 
We note that all signal points are located on the surface of an $N$-dimensional sphere of radius $\ro^{\text{cc}}=\sqrt{\E}$. 
However, there are multiple compositions that have the same sequence energy, and therefore, the surface is only partially utilized by the constant composition code as shown in Fig.~\ref{surface2sphere}.

There is no analytical expression for the distribution that maximizes the AIR for an ASK constellation.
For such constellations, Maxwell-Boltzmann distributions
\begin{equation}
P_{A}(a) = K\left(\lambda\right) e^{-\lambda a^2},\mbox{ $a \in \calA$}, \label{eq:mbdist}
\end{equation}
are pragmatically chosen for shaping amplitudes since they minimize the average energy for a given entropy, or equivalently, maximize the rate for a given average power constraint\footnote{Parameters $\lambda$ and $K(\lambda)$ determine the variance of the distribution and make sure that $\sum_{a\in\calA} P_{A}(a) = 1$.}~\cite{CoverT2006_ElementsofInfoTheo,frank1993}.
Moreover in~\cite{bocherer2015}, CCDM is included in the PAS scheme using MB distributions as the target $P_A$.
We stress that although these MB distributions maximize the energy efficiency, they do not maximize the AIR,
but operate very close to the capacity of ASK constellations~\cite[Table 5.1]{bochererbook}.

\subsection{Sphere Shaping}
We use sphere codes to represent the indirect approach to realize amplitude shaping.
The idea is to utilize sequences having an energy not larger than $\emax$.
Let $\Semax$ denote the set of bounded-energy $N$-amplitude sequences $s_1,s_2,\cdots,s_N$ that is defined as
\begin{eqnarray}
\Semax = \left\{ s_1,s_2,\cdots,s_N \bigg| \sum_{n=1}^{N} s_n^2 \leq \emax \right\}, \label{essset}
\end{eqnarray}
where $s_n \in \calA = \{1, 3,\cdots, 2^m-1\}$ for $n = 1, 2, \cdots, N$.
Throughout this paper, $\So$ will be used as the shorthand notation for this set.
The shaping rate of the sphere code which consists of all $N$-sequences in $\So$ is
\begin{eqnarray}
R_s = \frac{\log_2\left(\left|\So\right|\right)}{N}\mbox{   (bit/amp.).}
\end{eqnarray}
We note that $\So$ consist of all $2^m$-ASK lattice\footnote{We use ``$2^m$-ASK lattice'' to indicate the $N$-fold Cartesian product of $2^m$-ASK alphabet with itself.} points (the ones with positive components) on the surface of or inside the $N$-sphere of radius $\ro^{\text{sp}} = \sqrt{\emax}$ as shown in Fig.~\ref{surface2sphere}.
Similar to using MB distributions in constant composition shaping, we apply sphere shaping in a pragmatic manner to obtain high energy efficiency~\cite{laroia1994}, knowing that the AIR is not necessarily maximized.

\subsection{Comparison: Finite Length Rate Loss}
Let $\calC$ be a code which consists of amplitude sequences $a^N_k = a_{k1},a_{k2},\cdots,a_{kN}$ for $k=1, 2,\cdots, K$, where $K$ is the number of codewords of length $N$ in $\calC$. 
All codewords now occur with probability $1/K$.
Then we can write

\begin{align}
\log_2 K &= \ent(A_1,A_2,\cdots,A_N), \nonumber \\
	      &\leq  \sum_{n=1}^N \ent(A_n) \leq N\ent(A) \leq N\ent(A_{\text{MB}}),
\end{align}
where $A$ is a random variable with distribution $P(A=a) = \frac{1}{N} \sum_{n=1}^N P(A_n=a)$ for $a\in\calA$, and $A_{\text{MB}}$ is a MB-distributed random variable with expected symbol energy $\sum_{a\in\calA} a^2 P(A=a)$. 
This shows that the rate loss 
\begin{eqnarray}
\rloss \define \ent(A_{\text{MB}})-\frac{1}{N} \log_2 K,
\end{eqnarray}
is always non-negative.
As frequently done in the literature, see~\cite{ccdm,pbdm}, etc., we use the rate loss as a performance indicator for shaping codes at finite block lengths.

\begin{figure}[t]
\centering
\resizebox{\columnwidth}{!}{\includegraphics{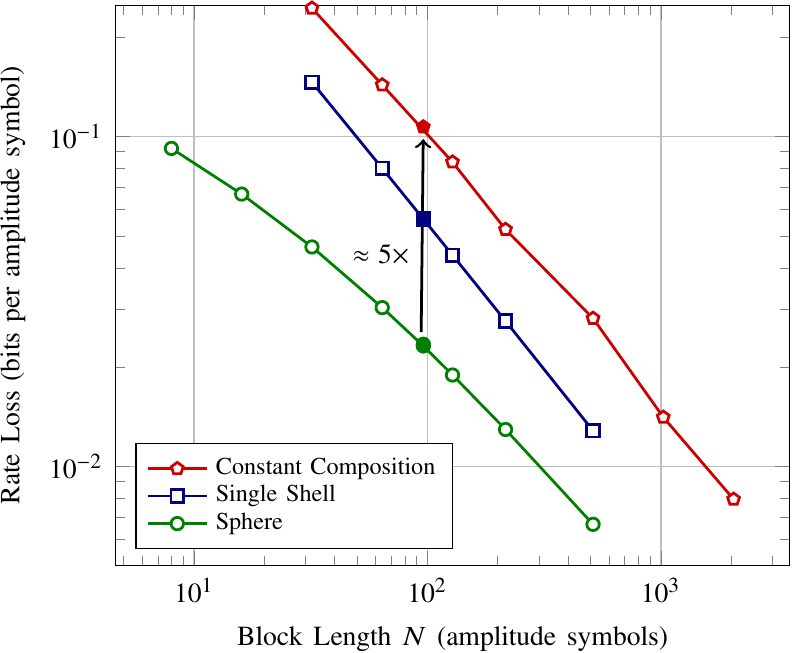}}
\caption{Comparison of constant composition, single-shell and sphere codes in terms of rate loss. 
The target shaping rate is $R_s = 1.75$~bit/amp. with 8-ASK.
}
\label{rloss}
\end{figure}

In Fig.~\ref{rloss}, we have compared sphere codes with constant composition codes for short block lengths in terms of rate loss.
For comparison, the rate loss is also plotted for single-shell codes which consist of all signal points on the surface of an $N$-dimensional sphere of radius $\ro^{\text{sh}}$ as shown in Fig.~\ref{surface2sphere}.
We fix the target shaping rate $R_s=1.75$~bit/amp. and
for each $N$, we choose $\#(a)$, $\ro^{\text{sh}}$ and $\emax$ for constant composition, single-shell and sphere codes, respectively, such that these codes contain at least $2^{NR_s}$ signal points.
We use the amplitude alphabet $\{+1,+3,+5,+7\}$ of 8-ASK.
{At $N=96$, the constant composition code experiences almost 5 times larger rate loss than the sphere code.
We note that 96 is the number of real dimensions used for data transmission in a single orthogonal frequency-division multiplexing (OFDM) symbol for one of the modes in the IEEE 802.11 standard~\cite[Table 17-5]{80211-2016}.}
It is no surprise that sphere codes have a smaller rate loss than constant composition codes, and it is shown in~\cite{ycgwic2018} that sphere codes minimize the rate loss at a given block length for a fixed rate.

\subsection{Comparison: Shaping Gain}
From a different perspective, at a fixed shaping rate $R_s$, the average sequence energy  
\begin{eqnarray}
\E = \frac{1}{K}\sum_{k=1}^{K} \sum_{a\in\calA} \#\left(a\big|a^N_k\right) a^2,
\end{eqnarray}
of constant composition and sphere codes can also be compared.
The following example is provided to advocate sphere shaping for short block lengths using average energy computations.

\begin{example}[{\bf Running example}]
Here we compare sphere and constant composition codes at block length $N=96$. 
We again target a shaping rate of at least $R_s = 1.75$~bit/amp. using the 8-ASK alphabet.
The constant amplitude composition satisfying this rate is $\{ \#(1), \#(3), \#(5), \#(7) \} = \{37, 30, 19, 10\}$ which has $\E = 1272$. 
For the sphere code that satisfies the rate constraint, $\emax=1120$, $R_s = 1.7503$, and $\E=1097$. 
We define the shaping gain with respect to uniform signaling, i.e., the reduction in average energy thanks to shaping\footnote{In \eqref{gainexp}, we assume that the average symbol energy expression $(2^{2m}-1)/3$ for uniform $2^m$-ASK constellations works as a good approximation for non-integer $m$. The shaping rate $R_s$ is increased by one in \eqref{gainexp} to account for the signed combinations of $N$-sequences.}, 
\begin{eqnarray}
\Gs = 10\log_{10} \left( \frac{2^{2(R_s+1)}-1}{3\E/N} \right),\mbox{   (in dB)}. \label{gainexp}
\end{eqnarray}
At $N=96$ and the target rate of $R_s=1.75$, constant composition and sphere codes have $\Gs=0.47$ and $\Gs=1.11$, respectively.
We see that the latter is 0.64 dB more energy-efficient.

Also note that there is another mode in the IEEE 802.11 standard where there are $N=216$ real dimensions reserved for data in a single OFDM symbol~\cite[Table 19-6]{80211-2016}.
At this block length, constant composition coding achieves $R_s=1.75$~bit/amp. by using $\{ \#(1), \#(3), \#(5), \#(7)\} = \{89, 69, 40, 18\}$ with $\E=2592$. 
Sphere coding requires $\emax=2456$ with $\E=2432$. 
Here the advantage of sphere codes over constant composition codes drops to 0.28 dB.
\end{example}

\subsection{Conclusion}
Motivated by the rate loss and energy efficiency discussions, we conclude that sphere shaping is suitable for block lengths smaller than a couple of hundreds symbols.
In the following, we will discuss an enumerative approach to realize $N$-sphere shaping.

\section{Enumerative Sphere Shaping} \label{sec:ess}
\subsection{Lexicographical Ordering}
Enumerative sphere shaping starts from the assumption that the amplitude sequences (in a sphere) can be lexicographically ordered.  
A sequence $a^N=a_1,a_2,\cdots, a_N$ is ``larger'' than  $b^N=b_1,b_2,\cdots, b_N$ if there exists an integer $n$ such that $a_j=b_j$ for $1\leq j \leq n-1$ and $a_n>b_n$. 
Then we write $a^N > b^N$.
Now we can define the index 
\begin{equation}
i\left(a^N\right) \define \left|\left\{ b^N \in \So : a^N > b^N\right\} \right|.  
\end{equation}
The mapping from sequences in $\So$ to indices is one-to-one, therefore 
\begin{equation}
a^N\left(i\right) = a^N \mbox{ if } i\left(a^N\right)=i.     
\end{equation}
We can use this mapping for transforming a message (index) into a sequence, and vice versa.
A look-up table (LUT) could be used to store the mapping.

\begin{example}[{\bf LUT to realize sphere shaping}]
Consider the parameters $\calA = \{1, 3, 5, 7\}$, $N=4$ and $\emax = 28$.
In Table~\ref{seqlist}, we see the corresponding sequences lexicographically ordered and their index.
We note that the amplitude 7 never occurs in Table~\ref{seqlist} since $\Emax=28$ does not allow any sequence to have it.

\begin{table}[ht]
\begin{center}
\caption{}
\scalebox{1}{
\includegraphics{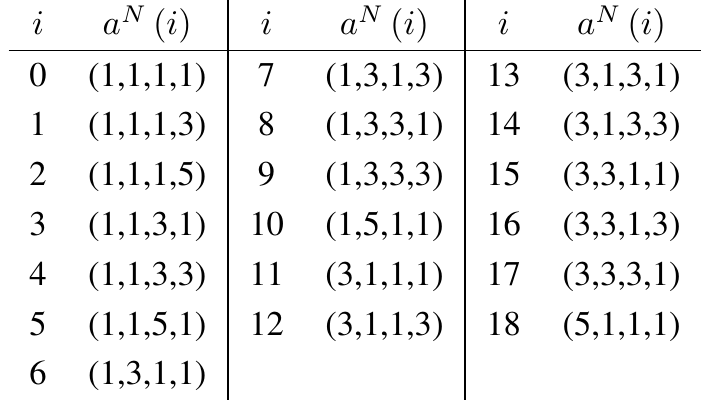}
}
\label{seqlist}
\end{center}
\end{table}
\end{example}

\subsection{Bounded-Energy Trellis}
\begin{figure}[t]
\centering
\resizebox{\columnwidth}{!}{\includegraphics{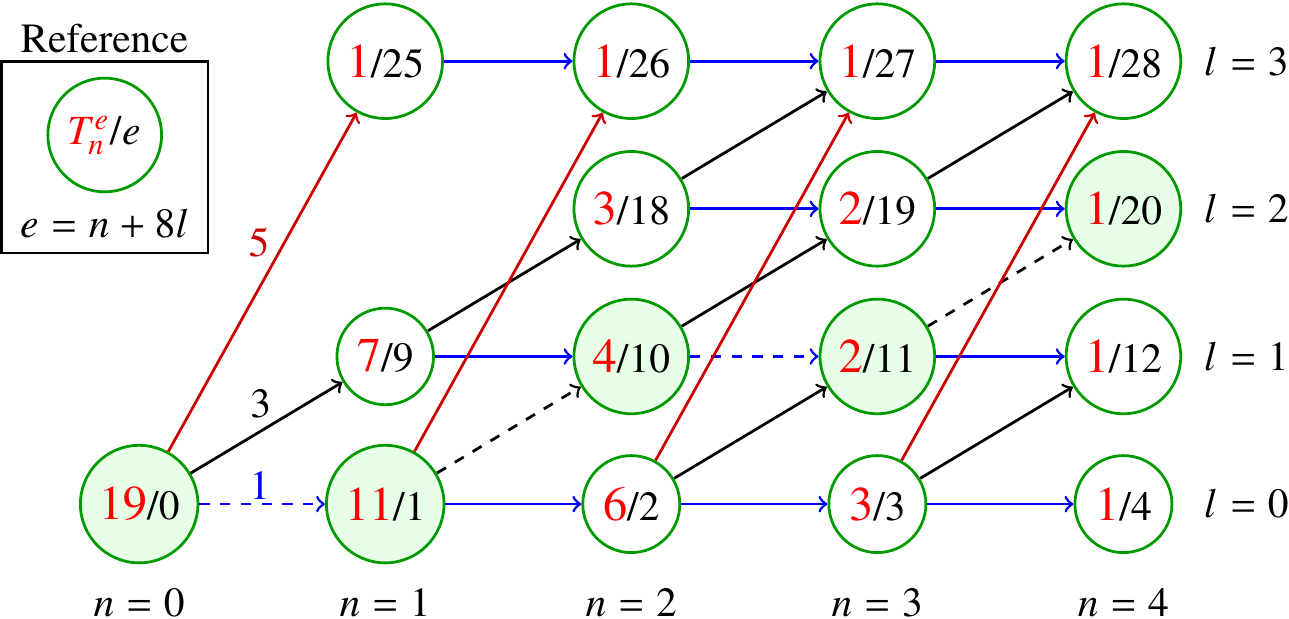}}
\caption{Bounded-energy enumerative amplitude trellis for $\calA = \{1, 3, 5, 7\}$, $N=4$ and $\emax=28$.}
\label{esstrellis}
\end{figure}

Using LUTs becomes impractical for realistic $N$, e.g., the number of entries is $2^{NR_s}= 2^{168}$ for $N=96$ and $R_s=1.75$ bit/amp.
Therefore, we need efficient methods to find the sequence corresponding to a message index, and vice versa.
For this purpose, we construct a bounded-energy trellis of which an example is given in Fig.~\ref{esstrellis} for the same parameters as in Table~\ref{seqlist}.

In this trellis, each sequence $a^N$ is represented by a unique path consisting of $N$ branches.
A branch connecting a node from column $n-1$ to column $n$ designates the $n^{\text{th}}$ component $a_n$ for $n = 1, 2,\cdots, N$.
The node that a path passes through in $n^{\text{th}}$ column identifies its accumulated energy over the first $n$ dimensions 
\begin{align}
e\left(a^n\right) \define \sum_{k=1}^{n} a_k^2.
\end{align}
These energy values are represented in small black letters in Fig.~\ref{esstrellis} for each node.
Each path starts from the zero-energy node (i.e., the bottom left) and ends in a node from the last column (i.e., at $n=N$).
The nodes in the last column are possible sequence energies $e(a^N)$ that take values from $\left\{ N, N+8, N+16,\cdots, \emax \right\}$ when amplitude sets in the form of $\calA = \{1, 3,\cdots\}$ are used.
Note that the energy $e$ of a node in column $n$ can be written as $e = n + 8l$ for $l = 0, 1,\cdots, L-1$ where 
\begin{eqnarray}
L = \lf \frac{\emax-N}{8} \rf +1, \label{eq:L}
\end{eqnarray}
is the number of energy levels in the final column.
Thus, a node can either be identified by the pair $(n,l)$ or $(n,e)$.
In the rest of the paper, we use the latter.

The larger red numbers in  Fig.~\ref{esstrellis} represent the number of paths that lead from their corresponding nodes to a final node. 
These numbers can be computed recursively for $n = N-1, N-2,\cdots, 0$ and $0 \leq e \leq E_{\max}$ as
\begin{equation}
T_n^e \define \sum_{a \in \mathcal{A}} T_{n+1}^{e+a^2}, \label{trellisCons}
\end{equation}
where the last column is initialized with ones, hence $T_N^e = 1$.
In column $n$, we only consider states with energy levels between $n$ and  $\emax +n -N$. 
We will show later how the trellis can be used to enumerate the sequences in a sphere. 
First, we give an example.

\begin{example}[{\bf Bounded-energy trellis}]
Consider node $(3,11)$ in Fig.~\ref{esstrellis}.
To go to a final node, the options are to use a branch of amplitude 1 or 3 which would lead the path to node $(4,12)$ or $(4,20)$, respectively.
Thus for the node $(3,11)$, $T_3^{11} = 2$.
Extending this reasoning, the number in the zero-energy node, $T_0^0 = 19$, is the number of sequences represented in this trellis i.e., $|\So|$.
We note that this number is consistent with Table~\ref{seqlist}.
\end{example}

The amplitude distribution $P_A$ of the sphere codebook over $\calA$ is
\begin{align}
P_A(a) = \frac{T_1^{a^2}}{\sum_{b\in\calA} T_1^{b^2}}.
\end{align}
The average energy of the sequences represented in the trellis is
\begin{align} 
\E  =  N \sum_{a\in\calA} P_{A}(a) a^2. \label{eq:Eformula}
\end{align}
For the amplitude trellis in Fig.~\ref{esstrellis}, the PMF is $P_A(a)=\{11/19, 7/19, 1/19, 0\}$ for $a \in \{1, 3, 5, 7\}$ which leads to $\E = 20.84$. 

\subsection{Enumerative Shaping and Deshaping}
The bounded-energy trellis can be used to compute the sequence with a given index, and vice versa.
Finding the index of a sequence $a^N$ is equivalent to count the number of sequences which are lexicographically smaller than $a^N$.
This can be implemented by considering the path representing $a^N$ in the trellis and adding the number of paths that branch off to lower nodes.
This leads to Cover's indexing formula~\cite{cover1973} for sequences in a sphere
\begin{equation}
i(a^N) = \sum_{n=1}^N \sum_{b<a_n} T_n^{b^2 + \sum_{j=1}^{n-1} a_j^2}. \label{deshaping}
\end{equation}
The following example based on the trellis in Fig.~\ref{esstrellis} demonstrates the indexing formula.

\begin{example}[{\bf Enumerative indexing}]
Consider $a^N = (1,3,1,3)$ which has the path passing through nodes $(0,0)$, $(1,1)$, $(2,10)$, $(3,11)$ and $(4,20)$, i.e., the path indicated by dashed lines and passing through the nodes filled with green in Fig.~\ref{esstrellis}.
At each transition for which there is a possible transition with a smaller amplitude, i.e., the second and the fourth transitions, we add the red numbers in the corresponding lower nodes, i.e., $T_2^2 = 6$ and $T_4^{12} = 1$, to find the total index which is 7.
This mapping is consistent with Table~\ref{seqlist}.
\end{example}

The indexing formula can be implemented in a recursive way, starting in the final node and following the path corresponding to the sequence to the root node, and accumulating the numbers of ``lower'' paths.
This is called deshaping and outlined in Algorithm~\ref{alg:essdeshaping}.
The inverse function, that determines from message index the sequence in a sphere, can also be implemented recursively. 
This inverse mapping is used for shaping and summarized in Algorithm~\ref{alg:essshaping}.

\begin{algorithm}[ht]
Given that the index satisfies $0 \leq i < T_0^0$, initialize the algorithm by setting the \textit{local index} $i_1 = i$. Then for $b\in\calA$ and $n= 1, 2,\cdots, N$:
    \begin{enumerate}
    \item 	
    Take $a_n=b$ be such that
      	\begin{equation}
		\sum_{b<a_n} T_{n}^{b^2+\sum_{j=1}^{n-1} a_j^2} \leq i_n < \sum_{b \leq a_n} T_{n}^{b^2+\sum_{j=1}^{n-1}a_j^2}, \label{shapCond}
      	\end{equation}
    \item 	
    and (for $n < N$)
      	\begin{equation}
	i_{n+1} = i_n - \sum_{b<a_n} T_{n}^{b+\sum_{j=1}^{n-1}a_j^2} . \label{shapUpd}
      	\end{equation}
    \end{enumerate} 
Finally output $a_1, a_2,\cdots, a_N$.   
\caption{Enumerative Shaping}
\label{alg:essshaping}
\end{algorithm}

\begin{algorithm}[ht]
  \caption{Enumerative Deshaping}
  Given $a_1, a_2,\cdots, a_N$:
  \begin{enumerate}
    \item    Initialize the algorithm by setting the \textit{local index} $i_{N+1} = 0$.
    \item    For $b\in\calA$ and $n = N, N-1, \cdots, 1$, update the local index as
      	\begin{equation}
      		i_n = \sum_{b<a_n} T_{n}^{b+\sum_{j=1}^{n-1}a_j^2} + i_{n+1}. \label{deshapUpd}
      	\end{equation}
   \item
     Finally output $i = i_1$.
  \end{enumerate}
  \label{alg:essdeshaping}
\end{algorithm}

\section{Implementation Aspects} \label{sec:implementation}
\subsection{Operational Rate and Unused Sequences}\label{ssec:unused}
Since we consider transmission of binary information, the number of multidimensional signal points that will actually be transmitted is an integer power of two.
Therefore, the input block length of a sphere shaping algorithm is defined as
\begin{eqnarray}
k = \left\lfloor\log_2\left(\left|\So\right|\right)\right\rfloor \mbox{ (bits)}. \label{iplen}
\end{eqnarray}
Thus, sequences having indices larger than or equal to $2^{k}$ are not utilized.
Depending on this set of omitted sequences, the operational average sequence energy changes.
Since ESS sorts sequences lexicographically, these unused sequences are not necessarily from the outermost shell, i.e., have the highest possible sequence energy.
Therefore, operational average energy can be larger but also smaller than $\E$ in~\eqref{eq:Eformula}. 
However, by deliberately removing branches from the enumerative trellis, $T_0^0$ can be decreased, i.e., the number of sequences can be made closer to an integer power of two.
Furthermore, we can make sure that the deleted sequences are from the outermost shell by removing connections to level $l=L-1$~\cite{GultekinW2019_OptimumTrellis}.

\begin{example}[{\bf Trellis modification}]
Consider the trellis in Fig.~\ref{esstrellis}. 
If the nodes $(1,25)$ and $(2,26)$, and the branches connected to them are removed, $T_0^0$ drops from 19 to 17.
The average sequence energy also decreases from $\E=20.84$ to 20.
\end{example}

The rate of the shaping algorithm is now $k/N$~bit/amp.
This rate can be easily adapted by tuning $\emax$, and thus, $k$. The granularity of this adaptation is $1/N$ which is the best possible.

\subsection{Required Storage and Computational Complexity}\label{complexity}
Enumerative shaping and deshaping algorithms require the storage of $T_n^e$ which needs a matrix of size $L$-by-$(N+1)$.
Each value of $T_n^e$ can be up to $\lceil NR_s \rceil$-bits long which upper bounds the required storage by $L(N+1)\lceil NR_s \rceil$~bits. 
These algorithms require at most $(|\calA|-1)$ subtractions or additions of $T_n^e$ values per dimension.
Thus, the computational complexity is at most $(|\calA|-1)\lceil NR_s \rceil$ bit operations per dimension\footnote{``Bit operation'' denotes one-bit addition or subtraction.} (bit oper./1-D) as shown in Table~\ref{tab:complexities}.

\begin{table}[ht]
\begin{center}
\caption{Full Precision Implementation}
\resizebox{\columnwidth}{!}{
\includegraphics{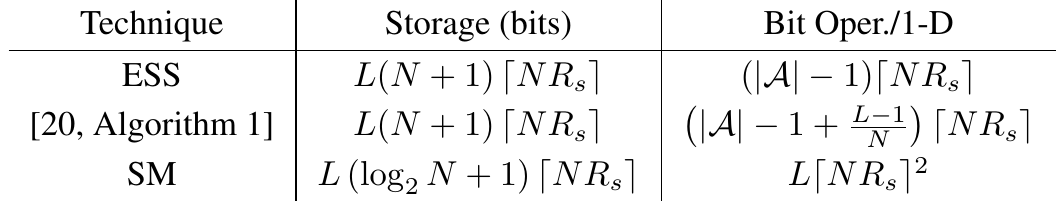}
}
\label{tab:complexities}
\end{center}
\end{table}

From the sphere-hardening result, we can conclude that $\emax \approx \E$ for large $N$. 
Approximating the required average energy to transmit $R$ bit/1-D. by $c2^{2R}$ for some constant $c$, we can write $\emax \approx Nc2^{2R}$~\cite[Sec. III.A]{laroia1994}.
Therefore $L$, see \eqref{eq:L}, can be considered as a linear function of $N$ for a given rate.
Thus, the required storage and computational complexity of ESS are cubic and linear in $N$, respectively.

\begin{example}[{\bf Running example}]
The storage of the trellis constructed with $N=96$ and $\emax=1120$ using the 8-ASK alphabet requires more than 264 kilobytes (kB) of memory.
Shaping and deshaping based on this trellis demand 507 bit operations per dimension. 
\end{example}

After discussing the energy efficiency and complexity of ESS for practical scenarios, we now summarize two different algorithms to realize sphere shaping, and compare them with ESS.

\subsection{Laroia's First Algorithm}\label{laroiafirst}
In~\cite{laroia1994}, Laroia \textit{et al.} provided two algorithms to realize sphere shaping, both of which sort the sequences based on their energy, i.e., based on the index of the $N$-dimensional shell that they are located on.
Sequences on the same shell are then ordered lexicographically.

To create algorithms that realize this ordering, a trellis can be constructed similar to Fig.~\ref{esstrellis}, now initializing the first column with only a single 1 in the lowest level, and then working rightwards~\cite[Sec. II-B]{laroia1994}. 
The numbers in the $n^{\text{th}}$ column and $i^{\text{th}}$ row of this trellis are the number of $n$-sequences of energy $e=n+8(i-1)$ for $n = 1, 2,\cdots, N$ and $i = 1, 2,\cdots, L$, see~\cite[(3)]{laroia1994}.
Consequently, the numbers in the last column of this trellis are the number of sequences located on the $i^{\text{th}}$ $N$-shell of the sphere for $i = 1, 2,\cdots, L$.

During shaping, where index-to-sequence mapping is realized,~\cite[Algorithm 1]{laroia1994} first finds the shell that the output sequence is located on.
This is realized by comparing the input index to the numbers in the last column of the trellis, and successively subtracting these numbers from the index whenever they are smaller.
This step requires at most $L-1$ comparisons and subtractions of at most $\lc NR_s \rc$-bit numbers per shaping operation.
After the $N$-shell to which the output sequence belongs is determined, the actual sequence is found by using~\cite[Algorithm 1]{laroia1994} explained in~\cite[Sec. II-C]{laroia1994}.
Thus, the required storage for this algorithm is similar to ESS, while its computational complexity is {slightly} higher due to the additional step explained above as shown in Table~\ref{tab:complexities}.
However, sequences are sorted based on their energy, and the unused sequences discussed in Sec.~\ref{ssec:unused} are from the outermost shell.

\subsection{Shell Mapping (Laroia's Second Algorithm)}
The second sphere shaping realization from~\cite{laroia1994} is based on the divide-and-conquer (D\&C) principle as used in~\cite{lang1989}.
This algorithm is known as shell mapping (SM) and demands the storage of $\log_2N+1$ columns of the shaping trellis instead of $N+1$.
Thus, the storage requirement is upper-bounded by $L(\log_2N+1)\lceil NR_s\rceil$~bits which behaves as $N^2\log N$ as a function of $N$.

Implementing SM requires up to $L$ full precision multiplications (or divisions) of at most $\lceil NR_s \rceil$-bit numbers at each step~\cite{ycgisit2018}.
Unlike ESS, the SM algorithm consists of $\log_2N$ steps.
However due to the nature of the D\&C principle, SM repeats the $n^{\text{th}}$ step $2^{n-1}$ times for $n=1, 2, \cdots, \log_2N$.
Expressing multiplications of $k$-bit numbers as $k^2$ bit operations as in~\cite{laroia1994}, the computational complexity of SM is upper-bounded by 
\begin{align}
\frac{1}{N} \sum_{n=1}^{\log_2N} 2^{n-1} L \lc NR_s \rc^2 &= \frac{1}{N}L\lc NR_s \rc^2 (N-1), \nonumber \\
&\leq L \lc NR_s \rc^2,
\end{align}
bit oper./1-D as shown in Table~\ref{tab:complexities}, which is cubic in $N$.

\begin{example}[{\bf SM complexity}]
To compare with our running example, the required number of bit operations per dimension is on the order of millions for SM at $N=96$.
Thus, we consider $N=32$ with $\emax=408$ which gives $k=56$, i.e., $R_s=1.7557$, and $\E=384$. 
Three parallel shell mappers can be used to shape over 96 dimensions in this case, however with a small loss of efficiency.
With these parameters, SM requires at most 155952 bit operations per dimension.
We conclude that SM is prohibitively complex for block lengths larger than a few dozens.
\end{example}

\subsection{Approximate Implementations}
To decrease the required storage for ESS and SM, we proposed a bounded-precision implementation in~\cite{ycgisit2018}.
Numbers in their corresponding trellises can be represented in base-2 as $T_n^e=m \cdot 2^p$ where $m$ and $p$ are called mantissa and exponent, stored using $n_m$ and $n_p$ bits, respectively.
Based on this representation, we modified the trellis computing equation \eqref{trellisCons} to
\begin{equation}
T_n^e \define \lf \sum_{a \in \mathcal{A}} T_{n+1}^{e+a^2} \rf_{n_m}, \label{trellisConsBP}
\end{equation}
where $\lfloor x \rfloor_{n_m}$ indicates rounding $x$ down to $n_m$ bits.
The result of \eqref{trellisConsBP} is then stored in the form $(m,p)$.
The trellis computed and stored using this idea is called the bounded-precision trellis.
We proved in~\cite{ycgisit2018} that the invertibility of ESS and SM operations is preserved for this bounded-precision approach.
We note that this can also be demonstrated for~\cite[Algorithm 1]{laroia1994} in a straightforward manner as well.
Obviously, since the numbers in a trellis decrease, this approximate way of implementing causes a rate loss.
However we showed that this rate loss is upper-bounded by $-\log_2(1-2^{1-n_m})$ bit/1-D~\cite{ycgisit2018}.

Required storage for ESS with this bounded-precision implementation is at most $L(N+1)(n_m+n_p)$~bits which is now quadratic in $N$.
Computational complexity per dimension of ESS now becomes at most $n_m(|\calA|-1)$ bit operations which is independent of $N$.
For SM, the memory demand drops to $L(\log_2N+1)(n_m+n_p)$~bits which behaves like $N\log N$.
The algorithm now requires at most $n_m^2L$ bit oper./1-D which is linear in $N$ as shown in Table~\ref{tab:bpcomplexities}. 

\begin{table}[ht]
\begin{center}
\caption{Bounded Precision Implementation}
\resizebox{\columnwidth}{!}{
\includegraphics{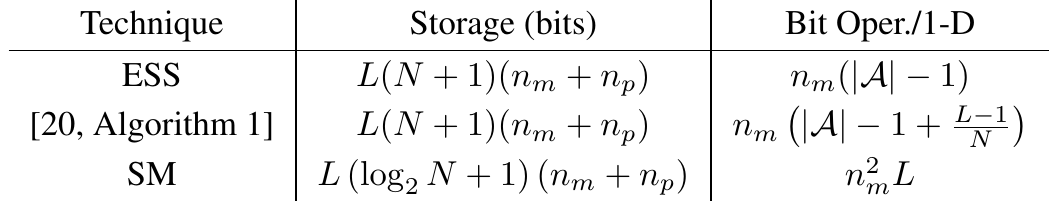}
}
\label{tab:bpcomplexities}
\end{center}
\end{table}

\begin{example}[{\bf Running example}]
If the ESS trellis of our running example at $N=96$ with $\emax=1120$ and $\calA=\{1, 3, 5, 7\}$ is now constructed using $n_m=12$~bit mantissas and $n_p=8$~bit exponents, the shaping rate drops from $R_s=1.7503$ to $1.7500$~bit/amp., still satisfying our target.
The input length does not change, i.e., $k=168$ for both full- and bounded-precision implementations, while the average energy increases from $\E=1096.9$ to $\E = 1097.1$. 
The storage requirement drops from 264 to 31.3 kB.
With the approximate implementation, the number of required bit operations per dimension also reduces from 507 to 36.
\end{example}

\subsection{Conclusion}
Thus, due to its rate loss performance over CCDM in the short block length regime, and significantly smaller computational complexity compared to SM, we propose to use ESS as the amplitude shaping method for block lengths smaller than a couple of hundreds. 
Laroia's first algorithm~\cite{laroia1994} is proposed around the same time as ESS~\cite{willems1993} and has similar performance with a slightly increased complexity. 

\begin{remark}[{\bf Complexity of AC-CCDM}]
We note here that the required storage and computational complexity of AC-based CCDM are minimal~\cite{ccdm}. 
In short, matching and dematching can be implemented by storing a couple of bytes, and with a few multiplications or divisions per input symbol, which makes CCDM perfectly suitable for large $N$ where its rate losses are insignificant. 
For more information on the algorithmic implementation and complexity of AC and AC-CCDM, we refer the reader to~\cite{ccdm},~\cite{ramabadran1990}, and~\cite[Ch. 5]{sayood2002lossless}.
\end{remark}

\section{Amplitude Shaping with Channel Coding} \label{sec:shapcode}
\subsection{Probabilistic Amplitude Shaping}
The concept of probabilistic amplitude shaping is introduced in~\cite{bocherer2015} to integrate amplitude shaping with existing binary FEC schemes.
The basic principle is to realize shaping over the amplitudes of the channel inputs and achieve error correction by coding the signs based on the binary labels of these amplitudes. 
Binary labeling of the $2^m$-ASK constellation points is assumed to be sign-symmetric, i.e., among the binary labels $B_1B_2\cdots B_m$ of a constellation point, $B_1$ selects its sign and $B_2B_3\cdots B_m$ specify its amplitude.
We call this {\it amplitude-shaped sign-coded modulation}.

\begin{figure*}[t]
\centering
\resizebox{1.45\columnwidth}{!}{\includegraphics{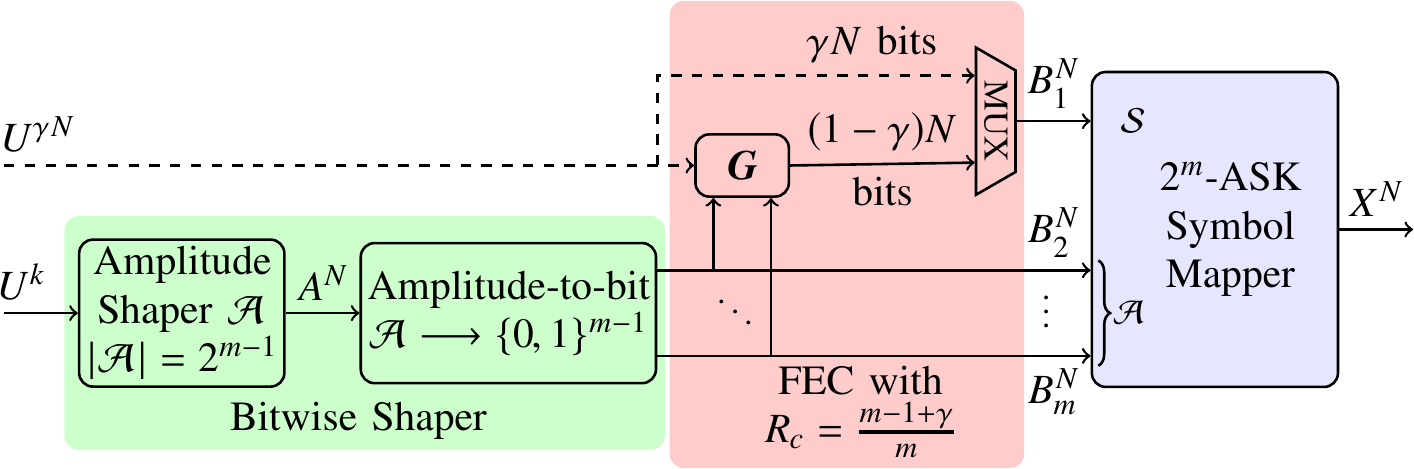}}
\caption{Block diagram of the probabilistic amplitude shaping transmitter. When $\gamma=0$, $(m-1)N$ amplitude bits are systematically encoded by a rate $R_c=(m-1)/m$ code to obtain $N$ sign bits.
When $\gamma>0$, $(m-1)N$ amplitude bits and $\gamma N$ additional (sign) bits are encoded by a rate $R_c=(m-1+\gamma)/m$ code to obtain $(1-\gamma)N$ sign bits.
In this case, the output of the encoder and the additional bits are used together as the signs.}
\label{TXRXblockdiag}
\end{figure*}

{We first consider the case where all sign bits are determined by the FEC encoder.}
At the transmitter, as shown in Fig.~\ref{TXRXblockdiag}, a $k$-bit uniform data sequence $U^k$ is mapped to a shaped amplitude sequence $A^N$ by applying the enumerative shaping function in Algorithm~\ref{alg:essshaping}.
Then these amplitudes are transformed into bits using the last $m-1$~bits of the labeling strategy. 
The binary label sequences $B^N_2 B^N_3\cdots B^N_m$ are used as the inputs to a rate $R_c = (m-1)/m$, systematic\footnote{A systematic code is a FEC code in which the input sequence is part of its output, to which parity bits are added.} FEC code which is specified by a $(m-1)N$-by-$mN$ generator matrix $\boldsymbol{G}$.
The code produces $N$ parity bits.
Finally, these parity bits are used to select the signs $S^N$ and the signed sequence $X^N = S^N A^N$ is transmitted over the channel.
The transmission rate of this structure is $R_t = k/N$~bit/1-D.

When a code of rate $R_c > (m-1)/m$ is employed, the amount of parity added by the encoder is not enough to select signs for all amplitudes.
Then an additional $\gamma N$-bit uniform data sequence $U^{\gamma N}$ is fed to a FEC code along with $B^N_2 B^N_3 \cdots B^N_m$.
This FEC code is specified by a $(m-1+\gamma)N$-by-$mN$ generator matrix $\boldsymbol{G}$.
Now the parity bits and extra data bits $U^{\gamma N}$ are used as the sign bit-level $B_1$.
Here $\gamma = mR_c-(m-1)$ denotes the fraction of signs that are selected by extra data.
In Fig.~\ref{TXRXblockdiag}, this setup has the dashed branches activated.
The transmission rate of this structure is $R_t = k/N+\gamma$~bit/1-D.

\subsection{Bit-Metric Decoding and Achievable Information Rate}
At the receiver, a soft demapper computes the log-likelihood ratio (LLR) of bit-level $B_i$ corresponding to the channel output $Y$ as
\begin{eqnarray}
L(B_i) = \log \left(\frac{ \sum_{x \in \calX_{i,0}} P_{X}(x)  f_{Y|X}\left(y|x\right) }{\sum_{x \in \calX_{i,1}} P_{X}(x)  f_{Y|X}\left(y|x\right)} \right), \label{eq:llr}
\end{eqnarray}
for $i=1, 2,\cdots, m$. Note that non-uniform a priori information on the symbols is taken into account.
Here $\calX_{i,u}$ indicates the set of $X \in \calX$ having $B_i=u$ in their binary label for $u \in \{0, 1\}$. 
Then a bit-metric decoder uses these LLR's to recover the transmitted data.
In~\cite{bocherer2014}, the rate
\begin{eqnarray}
\rbmd &=& \left[ \mathbb{H}(X) - \sum_{i=1}^{m} \mathbb{H}(\text{B}_i|Y ) \right]^+, \label{rbmd}
\end{eqnarray}
is shown to be achievable by a bit-metric decoder for any $P_{X}(x)$ where $[\cdot]^+ = \max\{0,\cdot\}$.

\subsection{Coding and Shaping Redundancy Balance} \label{redundancyshare}
\begin{figure}[t]
\centering
\resizebox{\columnwidth}{!}{\includegraphics{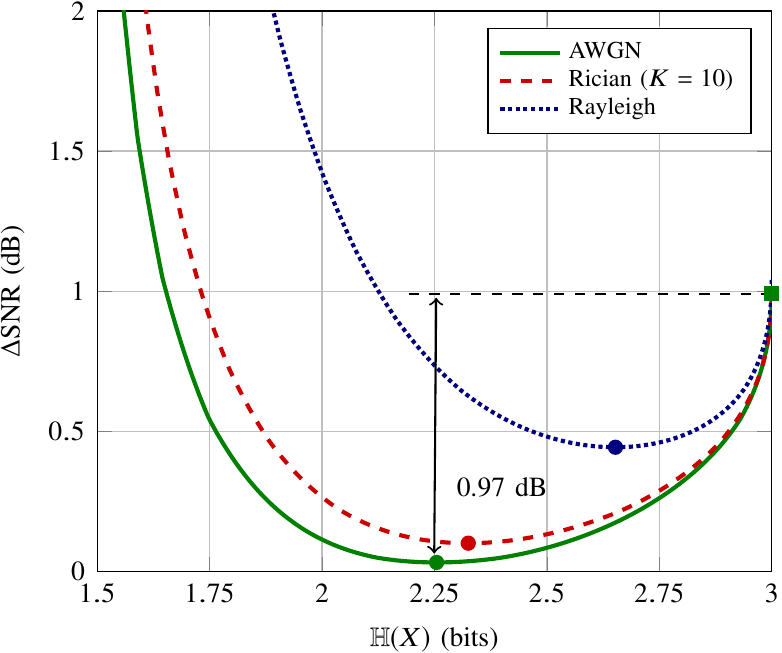}}
\caption{$\ent(X)$ vs. $\Delta\snr$ for 8-ASK at target rate $R_t = 1.5$. Filled circles specify the minima of their corresponding curves.}
\label{multiWP}
\end{figure}

In the PAS structure as in Fig.~\ref{TXRXblockdiag}, shaping and coding operations add $m-R_t$ bits redundancy per real dimension in total.
Amplitude shaping is responsible for $m-\ent(X)$ bits whereas FEC coding adds the remaining $\ent(X)-R_t$ bits.
Then assuming that $A$ is MB-distributed, for a fixed rate $R_t$ and constellation size $2^m$, the channel input entropy $\ent(X)=\ent(A)+1$ is a design parameter that can be used to tune the balance between the shaping and coding redundancies.
To find the optimum $P_A$ to achieve a fixed rate $R_t$, we use gap-to-capacity
\begin{align}
\delsnr = 10\log \left( \snr \right) \bigg|_{\rbmd=R_t} - 10\log \left( \snr \right) \bigg|_{\capc=R_t}, \label{gapeq}
\end{align}
as the metric that is to be minimized, similar to the approach followed by Wachsmann \textit{et al.} in~\cite{wachsmann1999}.
In Fig.~\ref{multiWP}, $\delsnr$ is plotted versus $\ent(X)$ at target rate $R_t = 1.5$ for the AWGN, Rayleigh and Rician channels\footnote{The fading parameter of the Rician distribution is set to $K=10$~\cite{goldsmithbook}.}.
Here, $A$ is assumed to be MB-distributed over the 8-ASK amplitude alphabet $\{1, 3, 5, 7\}${, i.e., $m=3$}.
For a variety of $\lambda$ values that keep $\ent(X) \in [1.5, 3]$, see \eqref{eq:mbdist}, $\rbmd$ is computed.
A binary reflected Gray code (BRGC) is used for labeling.
When computing $\delsnr$ for the AWGN and fading channels, the corresponding capacities are used as $C$ in \eqref{gapeq}.

The rightmost points of the curves in Fig.~\ref{multiWP} correspond to uniform, i.e., unshaped, signaling where the target rate is obtained by using a rate $R_t/m = 1/2$ FEC code.
Since there is no shaping, all $m-R_t=1.5$ redundant bits are added by the FEC encoder.
The leftmost parts of the curves correspond to uncoded signaling where the target rate is obtained by shaping the constellation such that $\ent(X)=R_t$.
Here all the redundancy is added by the shaper and the gap to capacity is infinite since error-free communication is only possible over a noiseless channel with uncoded signaling.
Remembering that the information rate is $R_t = \ent(A) + \gamma$ asymptotically as $N\rightarrow\infty$, and $\ent(X) = \ent(A) + 1$, the FEC code rate of PAS construction that corresponds to a point on the $\ent(X)$ vs. $\delsnr$ curve can be formulated as 
\begin{align}
R_c = \frac{m-1+\gamma}{m} 
    = \frac{m+R_t-\ent(X)}{m}. \label{eq:coderate}
\end{align}

The first deduction to be made from Fig.~\ref{multiWP} is that for a given constellation size $2^m$ and target rate $R_t$, there is an optimum balance between the shaping and coding redundancy in order to minimize the gap-to-capacity.
For example, under the AWGN channel assumption, the optimum point is roughly at $\ent(X)=2.25$ which prescribes that the shaping and coding operations should add 0.75 bits of redundancy each.
This, by (\ref{eq:coderate}), implies that a FEC code of rate $R_c=3/4$ should follow the shaper, i.e., the additional rate should be $\gamma = R_cm-(m-1)=0.25$.
At this point, the gain over uniform signaling is 0.97 dB\footnote{{We note here that it is not always possible to have a FEC code with the desired rate, especially in cases where existing codes are reused. In such cases, the FEC code rate which is closest to the optimum value should be employed. However, we argue that the loss due to this suboptimal parameter selection will be negligible since $\delsnr$ differs only slightly around the minimum in Fig.~\ref{multiWP}.}}.

The second conclusion is that as the channel becomes more and more dynamic, i.e., changes first to Rician and then to Rayleigh, (i) the optimum point shifts towards uniform signaling and the required coding redundancy increases, and (ii) the maximum capacity gain decreases.
For instance in the Rayleigh fading case, the optimum point is around $\ent(X)=2.65$ and thus the optimum FEC code rate is $R_c=0.62$. 
The corresponding capacity gain is 0.56 dB.
Therefore, we conclude that although the gains are smaller, shaping increases the maximum AIR over fading channels as well.
In addition, in such cases, the total redundancy should be distributed more in favor of coding and less in shaping.

\begin{figure*}[t]
\centering
\resizebox{1.45\columnwidth}{!}{\includegraphics{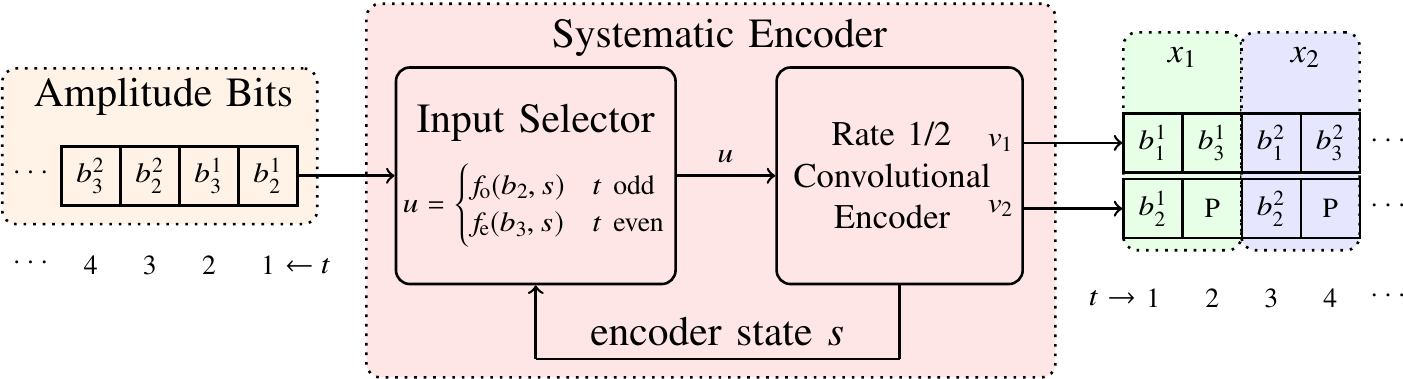}}
\caption{Block diagram of the PAS architecture employing the non-systematic convolutional code of the IEEE 802.11 standard~\cite{80211-2016}.
The encoder is preceded by the proposed input selector that realizes \eqref{eq:ipselect1} and \eqref{eq:ipselect2}.
These two combined (red box), operate as a systematic encoder, i.e., the input stream appears unchanged at the positions that corresponds to amplitude bits at the output.
{We note that since the FSM model of the encoder and its starting state are known to the input selector, the feedback link that carries the encoder state $s$ is unnecessary in practice, and it is only included here to emphasize that the selection process depends on $s$.}}
\label{inputselect}
\end{figure*}

\section{Employing the Non-Systematic Convolutional Code of IEEE 802.11 in PAS} \label{nsfecpas}
The PAS construction relies on the availability of a systematic FEC code.
In what follows, we  will explain how the non-systematic convolutional code (CC) applied in the IEEE 802.11 standard can be employed in the PAS framework together with an outer amplitude shaping block.

The mother CC used in the IEEE 802.11 standard has rate $R_c = 1/2$, and outputs the pair $(v_{1}[t],v_{2}[t])$ at time instance $t$ for the input bit $u[t]$.
Examining~\cite[Fig. 17-8]{80211-2016}, the output equations can be written as
\begin{align}
    v_{1}[t] &= u[t] \oplus u[t-2] \oplus u[t-3] \oplus u[t-5] \oplus u[t-6], \label{eq:op1} \\
    v_{2}[t] &= u[t] \oplus u[t-1] \oplus u[t-2] \oplus u[t-3] \oplus u[t-6]. \label{eq:op2}
\end{align}
The finite state machine (FSM) model of this code shows that in a given state, the output pair either belongs to the set $\{00,11\}$ or $\{01,10\}$ depending on the input bit $u[t]$ of the encoder.
Equivalently, by inverting the input symbol, the output pair will also be inverted.
This enables us to make half of the outputs equal to the values prescribed by the shaper, i.e., the amplitude bits.
The following example demonstrates this idea.

\begin{example}[{\bf Input selector}]
Consider the 8-ASK alphabet labeled with a BRGC and the CC with rate $R_c=2/3$, i.e., $\gamma=0$.
The puncturing pattern is $(1, 1, 1, 0)$.
As shown in Fig.~\ref{inputselect}, the output pairs of the encoder consist of a sign bit and an amplitude bit, i.e., $(B_1,B_2)$, for odd time indices $t$.
The pairs consist of an amplitude bit and a bit that will be punctured, i.e., $(B_3,P)$, for even time indices.
Our aim is to set half of the outputs to the prescribed amplitude bits, i.e., $v_2=b_2$ and $v_1=b_3$ for odd and even time indices, respectively.
From \eqref{eq:op2} and \eqref{eq:op1}, we get 
\begin{align}
    u[t] &= b_2 \oplus u[t-1] \oplus u[t-2] \oplus u[t-3] \oplus u[t-6] = f_{\text{o}}(b_2,s), \label{eq:ipselect1}
\end{align}
for odd $t$, and
\begin{align}
    u[t] &= b_3 \oplus u[t-2] \oplus u[t-3] \oplus u[t-5] \oplus u[t-6] = f_{\text{e}}(b_3,s), \label{eq:ipselect2}
\end{align}
for even $t$, respectively, where $s$ is the encoder state.
We call $f_{\text{o}}$ and $f_{\text{e}}$ the odd and even {\it input select functions}, respectively.
Using these functions, for each amplitude bit $b_2$ or $b_3$, the {\it input selector} in Fig.~\ref{inputselect} finds the input $u[t]$ to the convolutional encoder that will make the encoder output the prescribed amplitude bit in its corresponding position, i.e., on the $v_2$ branch for odd $t$, on the $v_1$ branch for even $t$.
The other output is determined either by \eqref{eq:op1} or \eqref{eq:op2}, and used as the sign bit or is punctured, respectively.
\end{example}

For other combinations of $m$ and $R_c$, the puncturing pattern and the positions of the amplitude bits at the output may change.
For such settings, the input select functions can be modified in a straightforward manner to ensure the encoder outputs the prescribed amplitude bits~\cite{ycgpimrc}.
{The block that realizes the inverse function of the input selector at the receiver can be implemented in a similar manner.
This block, given that the preceding FEC decoder correctly estimated the complete frame, does not introduce any errors.}

\begin{remark}[{\bf Non-systematic convolutional coding for PAS}]
For the sake of simplicity, we omitted the interleaver which is used for convolutionally coded transmission in the IEEE 802.11 standard~\cite{80211-2016}, but it can easily be included in the design of the proposed input selector~\cite{ycgpimrc}.
Furthermore, we note here that trellis termination can also be employed during encoding as specified in~\cite{80211-2016} by allowing a negligible decrease in shaping gain.
In this case, a couple of symbols at the end of the frame stay uniform due to zero padding which terminates the trellis.
Finally, a systematic, rate $k/n$ convolutional encoder is usually defined to have $k$ of its $n$ output branches reserved for systematic bits.
Since in our proposal the systematic bits may appear in different branches during encoding, the effective systematic encoder in Fig.~\ref{inputselect} (red box) differs from the common systematic encoders in the literature in general~\cite{JohanessonZ2015_FundamConvCod,Bossert1999_CodingTelecom}.
\end{remark}

\section{Numeric Results}\label{results}
In this section, we provide the Monte Carlo simulation results which are performed to determine the performance of ESS as the amplitude shaping method in the PAS scheme as shown in Fig.~\ref{TXRXblockdiag}.
We use Algorithm~\ref{alg:essshaping} and~\ref{alg:essdeshaping} as the amplitude shaping and deshaping functions.
A BRGC is applied by the symbol mapper.
The same mapping is used to label amplitudes at the output of the shaper.
The soft-demapper at the receiver computes LLRs using \eqref{eq:llr}.

\begin{table*}[ht]
\renewcommand{\arraystretch}{1.3}
\centering
\caption{}
\includegraphics{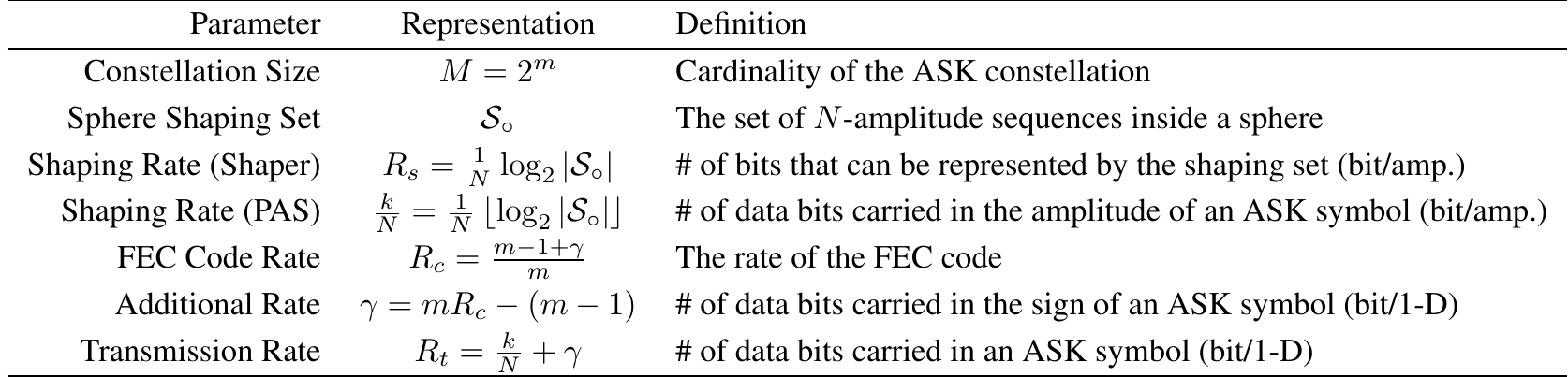}                             
\label{tab:parameters}
\end{table*}

\subsection{Simulation Settings}
Our objective is to compare uniform and shaped signaling structures fairly, i.e., at a fixed transmission rate $R_t$.
Shaping decreases the entropy $\ent(X)$ of the transmit constellation, and consequently the transmission rate.
Therefore, to compensate for this effect and operate at the same transmission rate $R_t$ as the compared uniform signaling scheme, either the constellation size $M$ and/or the FEC code rate $R_c$ of the PAS structure must be higher.

\begin{example}[{\bf Fair comparison}]
Consider a uniform signaling system which employs a rate $R_c=3/4$~FEC code followed by an 8-ASK symbol mapper.
The transmission rate of this system is $R_t=mR_c=2.25$~bit/1-D.
To obtain the same transmission rate with PAS, one of the following approaches can be taken.
First, the constellation size can be increased to 16-ASK while keeping the FEC code rate fixed at 3/4, which leads to $\gamma = mR_c-(m-1)=0$. 
Then the parameters of the amplitude shaper can be adjusted such that $k/N=2.25<m-1$ and consequently, $R_t=k/N+\gamma=2.25$.
Second, the FEC code rate can be increased to 5/6 while keeping the constellation size fixed at $M=8$, which leads to $\gamma=mR_c-(m-1)=0.5$.
Then the parameters of the amplitude shaper can be selected such that $k/N=1.75<m-1$ and consequently, $R_t=k/N+\gamma=2.25$.
\end{example}

As the FEC code, both non-systematic convolutional codes and systematic LDPC codes of rate $R_c$ and length $n_c$~code-bits are used as described in the IEEE 802.11 standard.
Each transmitted frame consists of $n_c$~coded bits, or equivalently, $N_c = n_c/m$~real symbols from the $2^m$-ASK alphabet.
Note that $N_c$ denotes the total number of real dimensions per FEC codeword.
In LDPC coding, we use the codeword lengths $n_c \in \{648, 1296\}$, and 50 iterations are performed by the decoder at the receiver.

Shaping is realized over $N$ real dimensions.
Note that when $N_c=N$, shaping and coding block lengths are the same.  
When on the other hand $N_c=\alpha N$ for some integer $\alpha>1$, each FEC frame consists of $\alpha$ shaped sequences.
At each target rate $R_t$ and constellation size $M=2^m$, we choose a FEC code rate $R_c$ for PAS based on the discussion in Sec.~\ref{redundancyshare}, i.e., Wachsmann curves. 
This FEC code rate results in $\gamma = mR_c - (m-1)$.
Then for ESS, $\emax$ is selected as the smallest value that satisfies $k/N + \gamma \geq R_t$.
For CCDM, the most energy-efficient composition $\{\#(a), a\in\calA\}$ that has at least $2^{k}$ sequences is selected.

Frequency-selective fading realizations are produced using type-D HiperLAN/2 channel model which is based on a Rician-modeled tapped delay line~\cite{hiperlan2}.
Doppler spread is taken to be zero.
Perfect channel state information is assumed to be available at the receiver.
For simulations over fading channels, OFDM is used as the modulation format as specified in the IEEE 802.11 standard.
The bandwidth is set to 40 MHz and separated into 128 subcarriers among which 108 are used for data, 6 are occupied by pilots and the remaining 14 are empty, see~\cite[Sec. 21.3.7.2]{80211-2016} for the actual subcarrier mapping\footnote{Prior to subcarrier mapping, two real ASK symbols are mapped to one complex quadrature amplitude modulation symbol.}.
The cyclic prefix length is taken to be 25 \% of an OFDM symbol duration.
For simulations over the AWGN channel, OFDM is omitted.
For simulations based on the convolutional codes, interleaving and trellis termination are realized as defined in~\cite{80211-2016}.

As the constellation, 4-ASK and 8-ASK are used during the simulations over fading channels which lead to 648 and 432 real dimensions per LDPC codeword of length $n_c=1296$~bits, respectively.
Thus a codeword (i.e., a frame), consists of three or two OFDM symbols for schemes based on 4-ASK and 8-ASK, respectively.
Shaping is realized over an OFDM symbol.

\subsection{Performance over the AWGN Channel}
\subsubsection{Convolutional Codes}
\begin{figure*}[t]
\centering
\resizebox{\columnwidth}{!}{\includegraphics{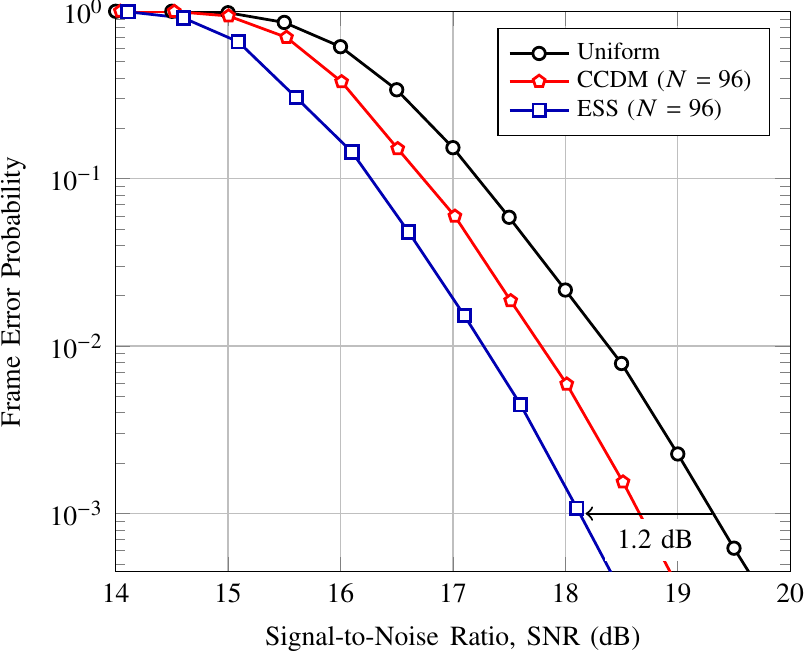}}
\resizebox{\columnwidth}{!}{\includegraphics{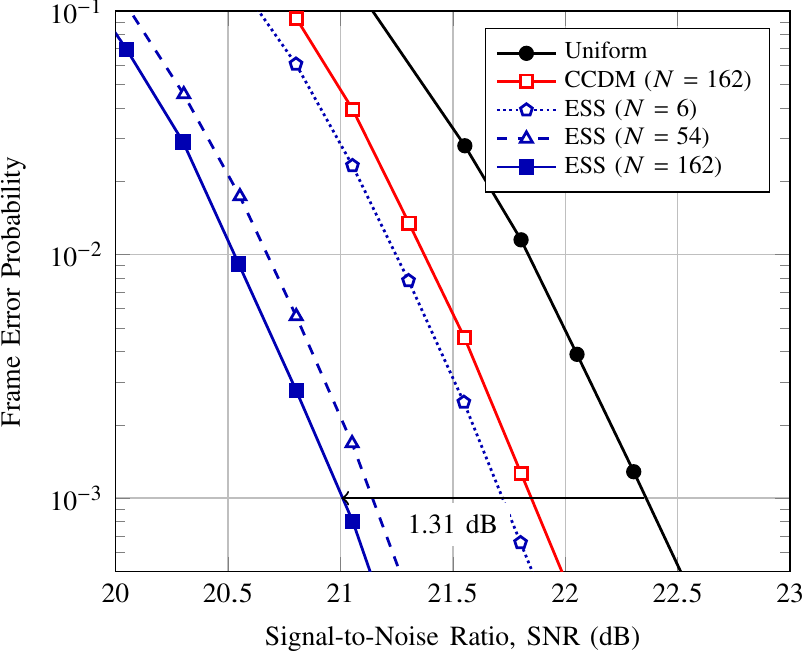}}
\caption{FER vs. SNR for ESS, CCDM and uniform signaling. ESS and CCDM are realized over $N$ dimensions, and combined with $R_c = 5/6$ FEC code. Uniform signaling is combined with $R_c = 3/4$ FEC code. (Left) Target rate is $R_t = 2.25$~bit/1-D. Convolutional coding and 8-ASK symbol mapping are used. (Right) Target rate is $R_t = 3$~bit/1-D. LDPC coding and 16-ASK symbol mapping are used.}
\label{CCperf}
\end{figure*}

The frame error rate (FER) performance of shaped and uniform signaling employing the non-systematic convolutional codes of the IEEE 802.11 standard over the AWGN channel can be found in Fig.~\ref{CCperf} (left).
Uniform signaling combines 8-ASK with the code of rate $R_c = 3/4$, leading to the target information rate $R_t = 2.25$~bit/1-D.
The FEC code rate $R_c$ that minimizes $\delsnr$ in \eqref{gapeq} for 8-ASK is approximately $5/6$ which should be combined with the MB distribution that gives $\ent(X)=2.75$.
Thus, ESS and CCDM are combined with the code of rate $R_c=5/6$ where $\gamma=1/2.$
Shaping is realized over $N=96$~dimensions using the parameters from our running example leading to $k/N=1.75$ for both methods.
We take 8 shaping blocks inside a single codeword which consists of $n_c=2304$~bits.
The shaping gain \eqref{gainexp} turns out to be $\Gs=1.11$~dB for ESS. 
We observe from Fig.~\ref{CCperf} (left) that ESS of 8-ASK is 1.2 dB more energy-efficient than uniform signaling at an FER of $10^{-3}$.
This roughly matches with the computed shaping gain.
In this setting, ESS outperforms CCDM by 0.55 dB.

\begin{table}[ht]
\centering
\caption{Parameters for ESS of 16-ASK for $R_t = 3$ ($R_c=5/6$, $\gamma=1/3$).}
\includegraphics{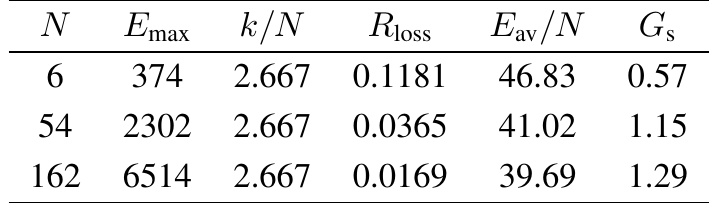}                               
\label{ldpcparameters}
\end{table}

\subsubsection{LDPC Codes}
The FER performance of shaped and uniform 16-ASK, combined with the systematic LDPC codes of length $n_c=648$~bits from the IEEE 802.11 standard over the AWGN channel is shown in Fig.~\ref{CCperf} (right).
Uniform signaling is combined with the code of rate $R_c=3/4$ leading to the information rate $R_t = 3$~bit/1-D.
For this constellation size and total rate, the FEC code rate that minimizes $\delsnr$ in \eqref{gapeq} is $0.85$ which should be combined with a MB distribution that gives $\ent(X)=3.6$.
Therefore, we combine ESS and CCDM with a code of rate $R_c=5/6$ which is the closest to $0.85$ in~\cite{80211-2016}, and $\gamma=1/3$.
Shaping is realized over $N \in \{ 6, 54, 162\}$~dimensions for ESS leading to $\{27, 3, 1\}$~shaping blocks inside a single codeword.
In Table~\ref{ldpcparameters}, corresponding parameters and metrics for ESS are tabulated.

We observe from Fig.~\ref{CCperf} (right) that at an FER of 10$^{-3}$ and at $N=6$, 54, and 162, ESS performs 0.59, 1.16 and 1.31 dB more energy-efficiently than uniform signaling, respectively.
These SNR improvements were quite well predicted by the shaping gains $\Gs$ from Table~\ref{ldpcparameters}.
Moreover, it is important to observe that ESS provides more than half a dB gain even at a very small dimensionality  $N=6$.
This enables a trade off between the complexity of ESS and the shaping gain $\Gs$.
Furthermore, when the primary objective is not to maximize the SNR gain but to provide a granular set of transmission rates, one can achieve this with ESS over only a couple of dimensions while still having a significant SNR improvement.
For comparison, MPDM~\cite{pbdm} needs at least 20 16-ASK symbols to only perform as good as uniform signaling where CCDM requires even more~\cite{fehenberger20182}.
Finally, we note that 95 \% of the gain achieved for block length $N=162$ can be reaped already for length $N=54$.

\begin{remark}[{\bf Shaping at $N=6$ using a LUT}]
{To realize sphere shaping at $N=6$ which is shown to provide 0.59 dB gain over uniform signaling in Fig.~\ref{CCperf} (right), a LUT which consists of $2^{k}=2^{16}$ entries is necessary. Considering that each entry includes 6 amplitudes, and amplitudes of 16-ASK can be represented with 3 bits, the size of the LUT is computed to be more than 147 kB. However, if the corresponding ESS trellis is computed with the bounded-precision approach presented in~\cite{ycgisit2018} using $n_m = 10$-bit mantissas and $n_p = 3$-bit exponents, only 4277 bits of memory is required for storage.}
\end{remark}
 
\begin{figure}[t]
\centering
\resizebox{\columnwidth}{!}{\includegraphics{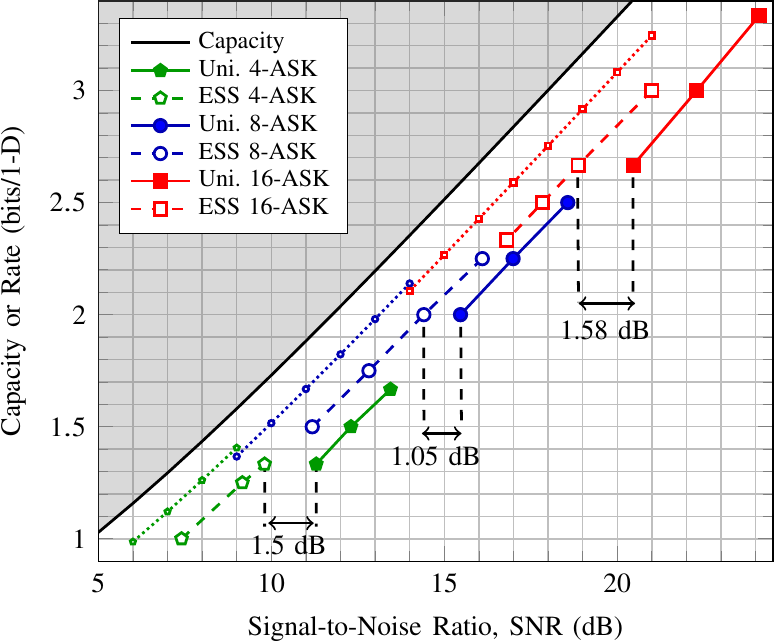}}
\caption{SNR values at which FER=10$^{-3}$ is achieved by ESS of 4-, 8- and 16-ASK. 
Shaped 8- and 16-ASK are combined with the $R_c=5/6$ LDPC code.
Shaped 4-ASK is followed by the $R_c=3/4$ code. 
The codes are of length $n_c=648$. 
This corresponds to $N=324$, $N=216$ and $N=162$ for 4-, 8- and 16-ASK, respectively.
Values for uniform signaling are also shown for $R_c \in \{2/3, 3/4, 5/6\}$.
{In~\cite[(1)]{polyanskiy2010finiteblocklength}, Polyanskiy {\it et al.} provided the normal approximation to the maximal achievable rate for the AWGN channel in the finite block length regime, more specifically, $C-\sqrt{V/N}Q^{-1}(\varepsilon)$, where $\varepsilon$ is the error probability, $Q^{-1}$ is the inverse $Q$-function, and $V$ is the AWGN channel dispersion as defined in~\cite[(293)]{polyanskiy2010finiteblocklength}.
Densely dotted red, blue and green curves show this approximation at $N=162$, $N=216$ and $N=324$ for $\varepsilon=10^{-3}$, respectively.}}
\label{capacityplot}
\end{figure}

In Fig.~\ref{capacityplot}, we plot the SNR values at which an FER of 10$^{-3}$ is obtained over the AWGN channel by ESS of 4-, 8-, and 16-ASK at different transmission rates.
In the same figure we also present the SNR values for uniform signaling with different FEC code rates.
ESS of 16- and 8-ASK employs a rate-5/6 code whereas ESS of 4-ASK is combined with a rate-3/4 code.
We note that $n_c=648$ corresponds to $N=$~324, 216 and 162 for 4-, 8- and 16-ASK, respectively.
{For comparison, we show the normal approximation to the maximal achievable rate for the AWGN channel at finite block lengths, i.e., $N \in \{162, 216, 324\}$, which was derived by Polyanskiy {\it et al.} in~\cite{polyanskiy2010finiteblocklength}.
We observe that for rates $R_t \in [1, 3]$~bit/1-D, it is possible to operate less than 1.5 dB away from this approximation.}
For instance at $R_t = 2.67$~bit/1-D, ESS performs 1.58 dB more efficiently than uniform signaling, and 1.4 dB away from the approximation at $N=162$.
Figure~\ref{capacityplot} can be applied to predict the performance of an ESS-shaped scheme at a given rate which may help the upper layers in the communication system selecting $R_t$ depending on the channel conditions.
It is unnecessary to apply FEC codes of different rates to provide rate granularity by moving this functionality to the shaping block and tuning $\emax$.

\subsection{Performance over Fading Channels}
\begin{figure}[t]
\centering
\resizebox{\columnwidth}{!}{\includegraphics{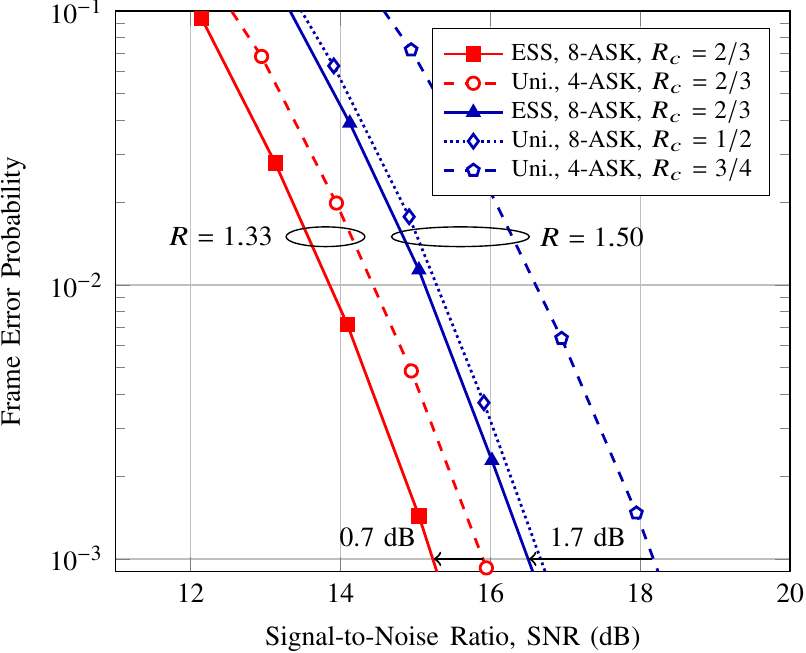}}
\caption{FER versus SNR behavior of ESS and uniform signaling over the HiperLAN/2-D channel. LDPC codes of length $n_c=1296$ are employed. Solid curves belong to ESS with $R_c=2/3$. Dashed and densely dotted curves belong to uniform signaling of $2^m$-ASK with FEC code rates $R_c \in \{1/2, 2/3, 3/4\}$. Shaping is over one OFDM symbol, i.e., $N=216$.}
\label{fadingfer}
\end{figure}

Figure~\ref{fadingfer} shows the SNR vs. FER for ESS and uniform signaling over a fading channel which is modeled by type-D HiperLAN/2~\cite{hiperlan2}.
Shaped results are based on 8-ASK where uniform curves correspond both to 8- and 4-ASK.
We argued in Sec.~\ref{redundancyshare} that as the channel starts to have a fading nature, the coding redundancy should increase relative to the shaping redundancy.
Therefore, we only use the smallest possible FEC code rate for the PAS scheme based on 8-ASK which is $R_c=2/3$. 

We see from Fig.~\ref{fadingfer} that at rate $R_t = 1.5$, ESS is 0.2 dB more efficient than uniform 8-ASK with $R_c=1/2$.
Furthermore, it outperforms uniform 4-ASK with $R_c=3/4$ by 1.7 dB\footnote{{In Fig.~\ref{fadingfer}, we provide FERs of two different uniform signaling settings at $R_t = 1.5$.
Among these, ESS provides 1.7 dB gain over the rate-3/4-coded uniform 4-ASK which is a combination that is supported by the IEEE 802.11 standard~\cite{80211-2016}.
On the other hand, the gain is 0.2 dB over the rate-1/2-coded uniform 8-ASK which is a combination that the IEEE 802.11 standard does not allow, but is simulated in this work for the sake of fairness.}}.
At rate $R_t = 1.33$, ESS requires 0.68 dB less SNR than uniform 4-ASK with $R_c=2/3$. 
{We note that the IEEE 802.11 standard does not provide any modulation order - coding rate combination that leads to $R_t = 1.33$~\cite{80211-2016}.}
The increase in gain here as $R_t$ decreases is due to the fact that the FEC code rate $R_c$ of the corresponding uniform setting also increases from $1/2$ to $2/3$, which degrades its performance with respect to the shaped scheme.
From Fig.~\ref{fadingfer}, we can conclude that shaping provides gains in fading scenarios. 
However, further study is required here.

\section{Conclusion}\label{sec:conc}
In this paper, we proposed enumerative sphere shaping (ESS) as the amplitude shaping method in the probabilistic amplitude shaping (PAS) scheme.
By employing all amplitude sequences satisfying a maximum-energy constraint, ESS makes use of the signal space in the most energy-efficient manner and thus, has lower rate loss than CCDM at any dimension.
The difference in rate loss (i.e., shaping rate minus Maxwell-Boltzman entropy) increases as the block length decreases, which makes ESS more suitable for applications with short data packets.

Sphere shaping can also be accomplished with shell mapping (SM), but we show here that ESS, and especially the approximated version of it, have significantly smaller computational requirements than SM. We should mention here that Laroia's first algorithm has performance similar to ESS with increased complexity.

The design of the communication system based on shaping in a PAS environment is extensively discussed with so-called Wachsmann curves as the starting point. For a given constellation size and target rate, these curves give us the shaping rate, coding rate and additional rate, that result in good performance. 

The superiority of ESS over other shaping methods for short block lengths makes ESS an interesting technique in wireless communication settings.   
In order to combine ESS with the non-systematic convolutional codes used in the IEEE 802.11 standard, we have developed a method here for realizing a PAS system. Shaping is done over single OFDM symbols and the decoder is the standard bit-metric decoder used in the 802.11 standard, that is modified only by using non-uniform a priori information in the bit-metrics.

Simulations over the AWGN channel demonstrate that ESS provides up to 1.6 dB improvement in SNR efficiency over uniform signaling for rates between 2-6 bits per complex channel use.
Finally, up to 0.7 dB gains are obtained over the frequency-selective channels modeled by HiperLAN/2 type-D.

\section*{Acknowledgment}
The authors would like to thank Dr. Semih \c{S}erbetli, Dr. Alex Alvarado and Dr. Tobias Fehenberger for fruitful discussions, and to the anonymous reviewers for their comments which helped to improve the presented paper.

\bibliographystyle{IEEEtran}
\bibliography{IEEEabrv,REFERENCES}

\end{document}